\documentclass[epj,twocolumn]{webofc}
\usepackage[varg]{txfonts}   % Web of Conferences font
\woctitle{The Innermost Regions of Relativistic Jets and Their Magnetic Fields}
\begin{document}
\title{Relativistic stellar jets: dynamics and non-thermal radiation}

\subtitle{Estimates on the dynamical impact of the stellar wind in low- and high-mass microquasars}

\author{Valent\'i Bosch-Ramon\inst{1}\fnsep\thanks{\email{vbosch@am.ub.es}}
        % etc.
}

\institute{Departament d'Astronomia i Meteorologia, Institut de Ci\`encies del Cosmos (ICC), Universitat de Barcelona (IEEC-UB), Mart\'i i
Franqu\`es 1, 08028 Barcelona, Spain}

\abstract{%
Relativistic stellar jets, produced in binary systems called microquasars, propagate through media with different spatial scales releasing their energy in the form of work and radiation from radio to gamma rays. There are several medium-interaction scenarios that these jets can face. In particular, in relativistic stellar jets the presence of a star is an unavoidable element whose importance deserves to be studied. In the case of high-mass stars, their powerful winds are likely to interact dynamically with the jet, but also low-mass stars in the post-main sequence phase can present dense winds that will act as an obstacle for the jet propagation. In this work, we present a semi-qualitative discussion on the importance of the star for the evolution of relativistic stellar jets.
}
\maketitle
\section{Introduction}
\label{intro}

Relativistic stellar jets are produced in binary systems called microquasars that consist of a a compact object and a companion star.
As their extragalactic relatives, relativistic stellar jets are to cover a long way through a complex medium, not only inhomogeneous in density but also with non-negligible velocities. As shown in a series of papers on numerical studies of relativistic stellar jets \cite{per08,bor09,per10,bos11,per12}, the stellar winds and the external medium on large spatial scales are expected to have a relevant dynamical role in the propagation and termination of relativistic stellar jets, as well as on their radiative output. 

In this paper, first we briefly discuss the physical scenario of the sources that host stellar relativistic jets. Then, we study in a semi-qualitative manner the main effects of the medium on the innermost regions of these jets.  Both types of microquasars, those hosting a high- and a low-mass star, are considered. 

This is the first time (to our knowledge) that the winds of low-mass stars are discussed as relevant dynamical obstacles for jet propagation on the spatial scale of the binary system.

\section{Microquasars}

{\it Microquasar} is a term broadly adopted in the nineties and later on to designate a class of X-ray binaries presenting radio jets, applied first to 1E 2740$-$2942 and GRS~1915$+$105 due to morphological and kinematical similarities with active galactic nuclei (AGN).  Namely, microquasars are binary systems that consist of a compact object, a neutron star or a black hole, which accretes material expelled by the stellar companion. The companion, as mentioned, can be a low- or a high-mass star. 
The combination of accretion, magnetic fields, and possibly the black hole rotation, gives rise to the bipolar jets characteristic of microquasars. These jets are typically seen in radio, and appear to share the same basic physics with those found in radio-loud AGN. For a relatively early but thorough classical review on microquasars we refer to \citep{mir99}; interestingly, some microquasars had been known as radio emitters for rather long, as it is the case of SS~433, whose extended radio jets were detected in 1979 \citep{spe79}.

As already mentioned in Sect.~\ref{intro}, several works have been devoted to study numerically the dynamical impact of the external medium on the microquasar jet. In particular, some of these works \cite{per08,per10,per12} studied the jet-medium interaction on the scales of the binary system in high-mass microquasars. This case is particularly clear in what refers to the degree of medium influence, given the powerful winds of massive stars. It is however less obvious, and to our knowledge not considered in some detail so far, the influence on the jet exerted by the matter lost by a low-mass star in its post-main sequence phases, say a red giant. 

We provide in the following some context for both high- and low-mass microquasars.

\subsection{High-mass microquasars}

Although just a handful of high-mass microquasars are known, these sources are likely powerful high-energy emitters and certainly powerful non-thermal sources, as their radio \citep[e.g.][]{mar01} and hard X-ray/soft gamma-ray \citep[e.g.][]{lau11} activity shows. Actually, despite the relatively small number of high-mass microquasars, there are already two of them that present gamma-ray emission: Cygnus~X-3, which has been clearly detected in the GeV range \citep{tav09,abd09}, and Cygnus~X-1, which over the years has provided with a number of marginal detections in both the TeV \citep{alb07} and the GeV \cite{sab10,mal13} ranges.  
Given that non-thermal processes, and in particular those leading to gamma-ray production, require of rather extreme conditions, such as fast efficient particle acceleration and large energy budgets, high-mass microquasars are attractive objects from the point of view of physics.

In binaries hosting a massive star and an accreting compact object, accretion generally takes place through wind-capture. The stellar winds in high-mass stars are fast and dense, with mass-loss rates $\dot{M}\gtrsim 10^{-7}\,M_\odot$/yr and velocities $v\sim 2000$~km/s \citep{muj12}, so the expelled stellar material can be relevant as an obstacle for the jet given both its high densities and thrust. This is already indicated by simple arguments, which we will present below. These simple arguments are complemented of course by the numerical simulations referred above, which confirm the strong wind influence. Finally, it is noteworthy that some evidence of wind-jet interaction may have been already found \citep[e.g.][]{vil13}.

Note that the importance of the star is two-fold in high-mass microquasars, because in addition to the dynamical impact of its wind, one has to account for the strong radiation ultraviolet bath of the binary surroundings, relevant for non-thermal radiation processes (as shown for instance in \citep{ara09} and references therein).

\subsection{Low-mass microquasars}

One of the first microquasars to be discovered was GRO~J1655$-$40. This source presented superluminal motion very similar to the radio components or blobs detected in AGN jets \citep{mir94}. This binary has a low-mass star in a post-main sequence stage of its evolution. 
In fact, most of the known microquasars harbour a low-mass star, and as GRO~J1655$-$40, several of them also have evolved stellar objects. For instance, GRS~1915$+$105, GX~339$-$4, and XTE~J1550$-$564 have sub-giant or giant stars as stellar companions \citep{gre01,cha02,mun08}. In addition, it would seem reasonable that the brightest X-ray binaries were powered by giant companions, because in that case a mass transfer larger than those of main sequence stars is expected \citep[e.g.][]{web83,rev11}. Therefore, it seems rather natural that several already known bright microquasars have post-main sequence stars.

Unlike in high-mass systems, in low-mass microquasars the compact object accretion and jet activity are fueled through Roche-lobe overflow. Given the faint stellar emission, low-mass microquasars may seem more alike to AGN, which are basically just accretion-ejection sources. Nevertheless, the role of the companion may deserve further attention, in particular if it is in a phase of strong mass loss. 

Up to now, the role of the star in low-mass microquasars has been basically restricted to be a mass donor for matter accretion in the compact object. However, this does not need to be the case under the light of the following considerations. Giants present strong mass-loss rates, of $\sim 10^{-7}\,M_\odot$/yr, and more evolved stars show even stronger mass-loss rates when in the asymptotic giant branch. Although these post-main sequence winds are slow, of $\sim 10$~km/s, and so carry small momentum as compared to the jet, these winds may still be relevant from the point of view of the mass with which they fill the binary surroundings.

\section{The impact of the star in the jet environment}

We consider now when the medium can be relevant for the propagation of a jet in both low- and high-mass microquasars. This is going to be a semi-quantitative discussion with some simple but illustrative estimates. We address the reader to the available, already cited, numerical simulations for theoretical support that complements what is discussed here (at least in what concerns high-mass systems). 

After being launched, the jet crosses a region that is expected to be dynamically dominated by the accretion disc or emanations from it, say a corona or a heavier and slower outflow as those found in AGN \citep[e.g.][]{kin13}. Putting aside the difficulty of disentangling the jet and its boundary with this disc outflow, which is likely to be smooth, it is expected that the momentum and energy injection into the system is dominated by the jet. Therefore, either when crossing the system for the first time, or when interacting with the stellar wind in a sort of stationary fashion, we consider here that it is the jet what directly interacts with the stellar wind, even if the former is contaminated by material of accretion disc origin.

\subsection{The wind as an obstacle}

The jet has to face regions loaded with the stellar wind material when crossing the binary regardless of the type of the stellar companion: 

For typical red-giant parameters and assuming a jet length-to-radius ratio of 10, the mass accumulated in front of the jet is roughly 
\begin{equation}
M_{\rm wj}\sim 10^{22}\dot{M}_{-7}R_{\rm orb13}v_{7}^{-1}\,{\rm g}, 
\end{equation}
where $\dot{M}_7=(\dot{M}/10^{-7}\,M_\odot/{\rm yr})$ is the wind mass-loss rate, $R_{\rm orb13}=(R_{\rm orb}/10^{13}\,{\rm cm})$ the binary separation distance, and $v_7=(v/10^7\,{\rm cm/s})$ the wind velocity. 

For a typical high-mass star, the mass of the medium in front of the jet is 
\begin{equation}
M_{\rm wj}\sim 10^{21}\dot{M}_{-7}R_{\rm orb13}v_{8}^{-1}\,{\rm g}, 
\end{equation}
where $v_8=(v/10^8\,{\rm cm/s})$. We are neglecting here the effect of wind focusing due to the gravitational field of the compact object, the formation of accretion streams, or orbital motion.

Finally, let us estimate the jet mass inside the binary region:
\begin{equation}
M_{\rm j}\lesssim 4\times 10^{17}L_{\rm j36}R_{\rm orb13}(\Gamma_{\rm j2}-1)^{-1}\,{\rm g}\ll M_{\rm wj}\,, 
\end{equation}
where $L_{\rm j36}=(L_{\rm j}/10^{36}\,{\rm erg/s})$ is the jet power and $\Gamma_{\rm j2}=(\Gamma_{\rm j}/2)$ the jet Lorentz factor. 

The jet-to-wind mass ratio on the binary scales is then $\sim 10^{-4}$.
This comparison already shows that the jet head will have to sweep large amounts of matter when crossing the binary system, suffering strong deceleration and forming a strong shock in the stellar wind. Unless the jet is strongly magnetized or hot, it is also expected the formation of a reverse shock in the jet itself. Also, the jet may be disrupted when crossing its dense environment, or may be perturbed enough when propagating through the stellar wind that the development of strong instabilities may break it when the jet head is already far from the binary (this scenario was explored in \citep{per08}). 

The impact of the medium mass will likely overcome internal jet dissipation effects, such as internal shocks, during the initial phases after jet launching unless jet production is very irregular. It seems then likely that when the jet is just formed and its head crosses the regions surrounding the binary system or slightly farther, the radiation processes will be mostly sustained by the interaction with the environment.

Once the jet has left the binary system, the density in both the jet and the wind goes as $\propto 1/d^2$, with $d$ being roughly the distance to the star. This implies that a jet that were substantially affected within the binary by the medium should still be affected by it even if the medium density got diluted. Only very powerful ejections that crossed the system not noticing the stellar wind would remain unaffected all the way up to their termination regions.

\subsection{The stellar wind versus the jet: energy and momentum}

Since the stellar wind also carries momentum and energy into the system, one can compare their injection by the wind and the jet into the binary:

A strong momentum rate for the wind ($\dot{p}_{\rm w}$) impacting on the jet would lead to strong bending and its likely destruction. Comparing momentum rates through the jet section as seen from the star and the compact object, one can estimate that the jet will acquire a lateral momentum from the wind of the order of its own for
\begin{equation}
L_{\rm j}\sim \dot{p}_{\rm w}c\sim 3\times 10^{35}\dot{M}_{-7}v_8\,{\rm erg/s}.
\end{equation}
Under such conditions, the energy of the jet may still be large enough to create a hot bubble that would eventually expand leaving the system. This however may not be possible if the energy injected by the stellar wind is higher than that of the jet. This case would correspond to a jet power
\begin{equation}
L_{\rm j}\sim 3\times 10^{34}\dot{M}_{-7}v_8\,{\rm erg/s}.
\end{equation}
Nevertheless, even powerful jets, with $L_{\rm j}\sim 10^{36}\dot{M}_{-7}v_8$~erg/s, may suffer important dynamical effects by perturbations in pressure exerted by the wind representing a substantial fraction of the jet ram pressure. Such perturbations can develop non-linearly until affecting the whole jet dissipating its kinetic energy in the form of heat, shocks, turbulence, and non-thermal particles \citep[see also][]{per10}. As shown in \cite{per12}, wind inhomogeneity will enhance the jet disruption processes sharpening the perturbations. Note that mass-entrainment due to the wind-jet lateral interaction will make the jet slower and heavier. 

One can estimate the effect of the wind lateral interaction on jet bending through a bending angle prescription
\begin{equation}
\theta\sim \dot{P}_w/\dot{P}_j\sim 0.3\,\dot{M}_{-7}V_{w8}\,L_{36}^{-1}. 
\end{equation}
Such a bending should be followed by asymmetric recollimation shocks.

In the specific case of low-mass microquasars with evolved stars, the wind will tend to be weaker than in high-mass systems because of a lower velocity. However, even so, the wind momentum and energy fluxes seem to be still important enough to undertake specific studies of this issue. 

\section{Discussion}

Numerical studies on scales a bit larger than the binary system, including orbital motion and the complex wind structure in the system, are missing. Nevertheless, one can explore and try to foresee the evolution of the jet if it has been substantially affected by the stellar wind. 

After crossing the binary, it is expectable that a significant fraction of the jet kinetic energy\footnote{A magnetically dominated jet on these scales cannot be discarded. We nevertheless assume here that at the relevant distances ($\gtrsim 10^6\,G\,M_{\rm BH}/c^2$) the jet is already matter-dominated.} has been converted into internal energy of some sort, making the jet more prone to the influence of a dynamic environment. The density and pressure of the environment decrease outwards, and the medium-affected jet, likely trans-sonic at least in some regions, is bent and its injection point changes along the orbit. All this will lead to: a medium negative pressure gradient accelerating the jet flow away from the star; stronger jet bending due to the effect of Coriolis forces; jet precession due to orbital motion; and generation of shocks, instabilities, and thus further kinetic energy dissipation. Not only in the binary region, but also further up, say up to $v_{\rm j}T_{\rm orb}\lesssim 10^{16}T_{6}$~cm \footnote{This corresponds to arcsec angular distances at few kpc distances.} ($v_{\rm j}$ and $T_{\rm orb}$ are the jet speed and the orbital period, respectively), the jet should be still strongly affected by the dynamics of the stellar wind together with the orbital motion. The related angular scales can be probed with radio interferometry. 

Therefore, under the effects of the environment a microquasar jet can be partially disrupted and mass-load. On other hand, even under disruption and orbital motion and related effects, the jet can inflate a strongly asymmetric bubble driven outwards by the symmetry of the orbital plane plus the density and pressure gradients of the stellar surroundings. One can then conclude that, at milliarcsec and arcsec scales, the jet cannot be laminar nor ballistic, with plenty of candidate sites for energy dissipation and thereby particle acceleration. In this regard, bending and asymmetric recollimation shocks inside the system may already reprocess up to a fraction $\sim \theta$ of the jet luminosity.

The radiative importance of the interaction of the jet with the stellar wind in both low- and high-mass microquasar is determined by the fraction of jet reprocessed energy that will go to non-thermal particles. In addition, the emissivities and radiation-channel relative importance will be determined by the target field densities (matter, radiation and magnetic field) and the impact of adiabatic cooling. Both leptonic and hadronic processes may take place in microquasars, and the morphological, variability and spectral properties of the non-thermal emission can be as complex as the medium in which it is produced \citep[see e.g.][]{bos09}.

%\begin{figure}
%\centering
%\sidecaption
%\includegraphics[width=5cm,clip]{tiger}
%\label{fig-3}      
%\end{figure}

\begin{acknowledgement} 
The author wants to thank the organizers for their kind invitation and for organizing a very fruitful scientific meeting.
V.B-R. acknowledges support by DGI of the Spanish Ministerio de Econom\'{\i}a
y Competitividad (MINECO) under grants  AYA2010-21782-C03-01 and FPA2010-22056-C06-02.
V.B-R. acknowledges financial support from MINECO through a Ram\'on y
Cajal fellowship. This research has been
supported by the Marie Curie Career Integration Grant 321520.
\end{acknowledgement}

%
% BibTeX or Biber users please use (the style is already called in the class, ensure that the "woc.bst" style is in your local directory)
% \bibliography{name or your bibliography database}

\end{document}